\documentclass[
    a4paper            
  ]
  {jpconf}

\usepackage{epsfig}
\usepackage{tabularx}
\usepackage{graphicx}
\usepackage{amssymb,amsfonts,amsmath}

\begin{document}

\title{Gravitino dark matter with neutralino NLSP in the constrained NMSSM}


\author{Grigoris Panotopoulos}

\address{Departament de Fisica Teorica, Universitat de Valencia, E-46100, Burjassot, Spain, and 
Instituto de Fisica Corpuscular (IFIC), Universitat de Valencia-CSIC,
Edificio de Institutos de Paterna, Apt. 22085, E-46071, Valencia, Spain}

\ead{grigoris.panotopoulos@uv.es}

\newcommand{\postscript}[2]{\setlength{\epsfxsize}{#2\hsize}
   \centerline{\epsfbox{#1}}}
\newcommand{\mweak}{M_{\text{weak}}}
\newcommand{\mplanck}{M_{\text{Pl}}}
\newcommand{\mstar}{M_{*}}
\newcommand{\OmegaM}{\Omega_{\text{M}}}
\newcommand{\OmegaDM}{\Omega_{\text{DM}}}
\newcommand{\ifb}{\text{fb}^{-1}}
\newcommand{\ev}{\text{eV}}
\newcommand{\kev}{\text{keV}}
\newcommand{\mev}{\text{MeV}}
\newcommand{\gev}{\text{GeV}}
\newcommand{\tev}{\text{TeV}}
\newcommand{\pb}{\text{pb}}
\newcommand{\mb}{\text{mb}}
\newcommand{\cm}{\text{cm}}
\newcommand{\km}{\text{km}}
\newcommand{\g}{\text{g}}
\newcommand{\s}{\text{s}}
\newcommand{\yr}{\text{yr}}
\newcommand{\sr}{\text{sr}}
\newcommand{\eg}{{\em e.g.}}
\newcommand{\ie}{{\em i.e.}}
\newcommand{\ibid}{{\em ibid.}}
\newcommand{\eqsref}[2]{Eqs.~(\ref{#1}) and (\ref{#2})}
\newcommand{\secref}[1]{Sec.~\ref{sec:#1}}
\newcommand{\secsref}[2]{Secs.~\ref{sec:#1} and \ref{sec:#2}}
\newcommand{\figref}[1]{Fig.~\ref{fig:#1}}
\newcommand{\figsref}[2]{Figs.~\ref{fig:#1} and \ref{fig:#2}}
\newcommand{\sla}[1]{\not{\! #1}}

\newcommand{\mB}{m_{B^1}}
\newcommand{\mq}{m_{q^1}}
\newcommand{\mf}{m_{f^1}}
\newcommand{\mKK}{m_{KK}}
\newcommand{\WIMP}{\text{WIMP}}
\newcommand{\SWIMP}{\text{SWIMP}}
\newcommand{\NLSP}{\text{NLSP}}
\newcommand{\LSP}{\text{LSP}}
\newcommand{\mWIMP}{m_{\WIMP}}
\newcommand{\mSWIMP}{m_{\SWIMP}}
\newcommand{\mNLSP}{m_{\NLSP}}
\newcommand{\mchi}{m_{\chi}}
\newcommand{\mgravitino}{m_{\gravitino}}
\newcommand{\mmed}{M_{\text{med}}}
\newcommand{\gravitino}{\tilde{G}}
\newcommand{\Bino}{\tilde{B}}
\newcommand{\photino}{\tilde{\gamma}}
\newcommand{\stau}{\tilde{\tau}}
\newcommand{\snu}{\tilde{\nu}}
\newcommand{\squark}{\tilde{q}}
\newcommand{\epsEM}{\varepsilon_{\text{EM}}}

\newcommand{\rem}[1]{{}}

\def\tauLbar{\overline {\tau_L}}
\def\Real{\Re e\,}
\def\Imag{\Im m\,}
\def\mg{m_{\tilde{G}}}
\def\msl{m_{\tilde{l}}}
\def\pstau{p_{\tilde{l}}}
\def\ptau{p_{{l}}}
\def\pZ{p_V}
\def\pg{p_{\tilde{G}}}

\begin{abstract}
The gravitino dark matter with neutralino NLSP hypothesis is
investigated in the framework of NMSSM. We have considered both 
the thermal and non-thermal
gravitino production mechanisms, and we have taken into account all
the collider and cosmological constraints. The maximum allowed
reheating temperature after inflation, as well as the maximum
allowed gravitino mass are determined.
\end{abstract}


\section{Introduction}

There is accumulated evidence both from astrophysics
and cosmology that about 1/4 of the energy budget of the universe
consists of so called dark matter~\cite{Munoz:2003gx}, namely a 
component which is
non-relativistic and neither feels the electromagnetic nor the
strong interaction. It is known that in the standard model of
particle physics there is not a suitable dark matter candidate.
Supersymmetry, a very well-motivated idea, is perhaps the most
popular way beyond the standard model, and it provides us with an
ideal candidate for playing the role of cold dark matter in the
universe.

Out of the several dark matter candidates, certainly the most
well-studied case is the one of the neutralino. However, the
gravitino is also a very compelling candidate for several reasons.
Its interactions are completely fixed by the supergravity
Lagrangian, its mass $\mgravitino$ is directly related to the scale
at which supersymmetry is broken, and finally the reheating
temperature after inflation $T_R$, which is essential for the
precise baryon asymmetry generation mechanism, plays a certain role
in gravitino cosmology.

The MSSM~\cite{mssm} is the most economical supersymmetric extension of the
standarm model. However, it suffers from the so-called $\mu$
problem~\cite{Kim:1983dt}. The natural values for the higgs/higgsino 
mass parameter
$\mu$ is either $0$ or the Planck mass, while phenomenology requires
the $\mu$ should be at the electroweak scale. The simplest
supersymmetric model that solves the $\mu$ problem is the 
NMSSM~\cite{Nilles:1982dy},
where one more singlet chiral superfield is introduced.

It is therefore interesting to try to see whether a particular
cosmological scenario, with gravitino LSP and neutralino NLSP in the
framework of the NMSSM, is a viable one.

\section{The theoretical framework}

\subsection{Basics of NMSSM}

The particle physics model is defined by the superpotential
\begin{equation}\label{2:Wnmssm}
W= \epsilon_{ij} \left( Y_u \, H_2^j\, Q^i \, u + Y_d \, H_1^i\, Q^j
\, d + Y_e \, H_1^i\, L^j \, e \right) - \epsilon_{ij} \lambda \,S
\,H_1^i H_2^j +\frac{1}{3} \kappa S^3\,
\end{equation}
as well as the soft breaking masses and couplings
\begin{eqnarray}\label{2:Vsoft}
-\mathcal{L}_{\textrm{soft}}=&
 {m^2_{\tilde{Q}}} \, \tilde{Q}^* \, \tilde{Q}
+{m^2_{\tilde{U}}} \, \tilde{u}^* \, \tilde{u} +{m^2_{\tilde{D}}} \,
\tilde{d}^* \, \tilde{d} +{m^2_{\tilde{L}}} \, \tilde{L}^* \,
\tilde{L} +{m^2_{\tilde{E}}} \, \tilde{e}^* \, \tilde{e}
 \nonumber \\
& +m_{H_1}^2 \,H_1^*\,H_1 + m_{H_2}^2 \,H_2^* H_2 +
m_{S}^2 \,S^* S \nonumber \\
& +\epsilon_{ij}\, \left( A_u \, Y_u \, H_2^j \, \tilde{Q}^i \,
\tilde{u} + A_d \, Y_d \, H_1^i \, \tilde{Q}^j \, \tilde{d} + A_e \,
Y_e \, H_1^i \, \tilde{L}^j \, \tilde{e} + \textrm{H.c.}
\right) \nonumber \\
& + \left( -\epsilon_{ij} \lambda\, A_\lambda S H_1^i H_2^j +
\frac{1}{3} \kappa \,A_\kappa\,S^3 + \textrm{H.c.} \right)\nonumber \\
& - \frac{1}{2}\, \left(M_3\, \lambda_3\, \lambda_3+M_2\,
\lambda_2\, \lambda_2 +M_1\, \lambda_1 \, \lambda_1 + \textrm{H.c.}
\right) \,
\end{eqnarray}
When the singlet acquires a vaccum expectation value, S, we obtain an
effective $\mu$ parameter, $\mu_{eff}=\lambda S$. Imposing
universality at the GUT scale, a small controllable number
of free parameters remains, namely \\
\centerline{$tan \beta=v_u/v_d, m_0, A_0, m_{1/2}, \lambda, A_k$}
and the sign of the effective $\mu$ parameter can be chosen at will.

Because of the extra singlet superfield, in the NMSSM there is a
larger higgs sector and a larger neutralino sector.
The neutralino mass matrix is characterized by the appearence of
a fifth neutralino state, meaning that 
the composition of the lightest neutralino has an extra
singlino contribution
\begin{equation} \label{composition}
\tilde \chi^0_1 = N_{11} \tilde B^0 + N_{12} \tilde W_3^0 + N_{13}
\tilde H_1^0 + N_{14} \tilde H_2^0 + N_{15} \tilde S\,
\end{equation}
In the following, neutralinos with $N^2_{11}>0.9$, or
$N^2_{15}>0.9$, will be referred to as bino- or singlino-like,
respectively.

Furthermore, in the Higgs sector we have now two CP-odd neutral, and
three CP-even neutral Higgses. We make the assumption that there is
no CP-violation in the Higgs sector, and therefore the CP-even and
CP-odd states do not mix. We are not interested in the CP-odd
states, while the CP-even Higgs interaction and physical eigenstates
are related by the transformation
\begin{equation}\label{2:Smatrix}
h_a^0 = S_{ab} H^0_b\,
\end{equation}
where $S$ is the unitary matrix that diagonalises the CP-even
symmetric mass matrix, $a,b = 1,2,3$, and the physical eigenstates
are ordered as $m_{h_1^0} < m_{h_2^0} < m_{h_3^0}$.

\subsection{Production of gravitinos}

After inflation gravitinos can be produced in
two ways. One way to produce gravitinos is with scatterings from the
thermal bath, and another is from the out-of-equillibrium decays of
the NLSP, which decouple from the thermal bath before primordial
Big-Bang Nucleosynthesis and decay after the BBN time. Thus, imposing
the WMAP bounds~\cite{Komatsu:2008hk} we can write for the gravitino abundance
\begin{equation}
0.1097 < \Omega_{3/2}h^2=\Omega_{3/2}^{TP}h^2+Y_{3/2}^{NLSP}h^2 < 0.1165
\end{equation}
where
\begin{equation}
\Omega_{3/2}^{NLSP} h^2 = \frac{\mgravitino}{m_{NLSP}} \:
\Omega_{NLSP} h^2
\end{equation}
with $m_{NLSP}$ the mass of the NLSP, and $\Omega_{NLSP} h^2$ the
abundance the NLSP would have, had it not decayed into the
gravitino. The thermal contribution is given by (approximately for a
light gravitino, $\mgravitino \ll m_{\tilde{g}}$)~\cite{Bolz:2000fu}
\begin{equation}
\Omega_{3/2}^{TP} \simeq 0.27 \: \left ( \frac{T_R}{10^{10}~GeV}
\right ) \: \left ( \frac{m_{\tilde{g}}}{TeV} \right )^{-2} \: \left
( \frac{\mgravitino}{100~GeV} \right )
\end{equation}

\subsection{Neutralino life-time}

 The NSLP is unstable with a lifetime that is
typically larger than BBN time $t_{BBN} \sim 1$~sec. Energetic
particles produced by the NLSP decay may dissociate the background
nuclei and significantly affect the primordial abundances of light
elements. If such processes occur with sizable rates, the
predictions of the standard BBN scenario would be altered and the
success of the primordial nucleosynthesis would be spoiled. BBN
constraints on cosmological scenarios with exotic long-lived
particles predicted by physics beyond the Standard Model have been
studied, and here we have used the figure 2 of Ref.~\cite{Kawasaki:2004yh}. 
Thus, we need to compute the lifetime of the neutralino.
There are three main decay channels, namely neutralino decaying into
gravitino and a standard model particle that can be one of the
followings: Photon, Z boson and Higgs. The supergravity Lagrangian
is known, and the Feynman rules for the gravitino interactions have
been derived~\cite{Moroi:1995fs}. Therefore, the neutralino
lifetime is given by
\begin{eqnarray}
\tau & = & \frac{1}{\Gamma} \\
\Gamma & = & \Gamma(\chi \to \gamma \gravitino)+\Gamma(\chi \to Z
\gravitino)+\Gamma(\chi \to h \gravitino)
\end{eqnarray}
and the exact expressions for the diferent contributions can be found
in~\cite{Barenboim:2010mn}.

\section{Constraints and results}

We have used NMSSMTools~\cite{Ellwanger:2004xm}, a computer software
that computes the masses of the Higgses and the superpartners, the
couplings, and the relic density of the neutralino, for a given set
of the free parameters. We have performed a random scan over the whole
parameter space (with fixed $\mu > 0$ motivated by the muon
anomalous magnetic moment), and we have selected only those points
that satisfy i) theoretical requirements, such as neutralino LSP,
correct electroweak symmetry breaking, absence of tachyonic masses
etc., and ii) LEP bounds on the Higgs mass, collider bounds on SUSY
particle masses, and experimental data from B-physics.
Finally, for any given point in the cNMSSM
parameter space, the neutralino lifetime is a function of the
gravitino mass only. Imposing the BBN constraints we find the
maximum allowed gravitino mass, and from the cold dark matter bound
we can determine the maximum allowed reheating temperature. 
For all the acceptable points the lightest neutralino is either a bino or a
singlino, and our main results are summarized in the figures below.

 For the bino case, figure~1 shows the maximum allowed
reheating temperature after inflation versus the maximum allowed
gravitino mass, both in GeV. Although it cannot be seen directly
from the figures, the maximum possible gravitino mass in the bino
case is $\mgravitino \simeq 1$~GeV, and the corresponding reheating
temperature is $T_R \sim 10^7$~GeV. Therefore, we see that a) the
gravitino in this scenario must be much lighter than the rest of
superpatners, but nevertheless it is still in the gravity-mediated
SUSY-breaking scheme, and b) the reheating temperature after
inflation is not large enough for thermal leptogenesis.

For the singlino case, we show in figure~2 the maximum
allowed reheating temperature versus gravitino mass, both in GeV.
This time the neutralino relic density is even larger than before,
and gravitino now must be extremely light. This is due to the
smallness of the coefficients $N_{11}, N_{12}$ in the decay rate to
gravitino and photon. For the same lifetime as before, the gravitino
mass must be several orders of magnitude lower than in the bino
case. The last figure shows that in the singlino case the reheating
temperature cannot be larger than about $200$~GeV. However, this
value is much lower than the minimum value required for the
computation of the gravitino thermal production $T_R > 1$~TeV, and therefore we
conclude that this scenario must be excluded.

\begin{figure}
\begin{center}
\includegraphics{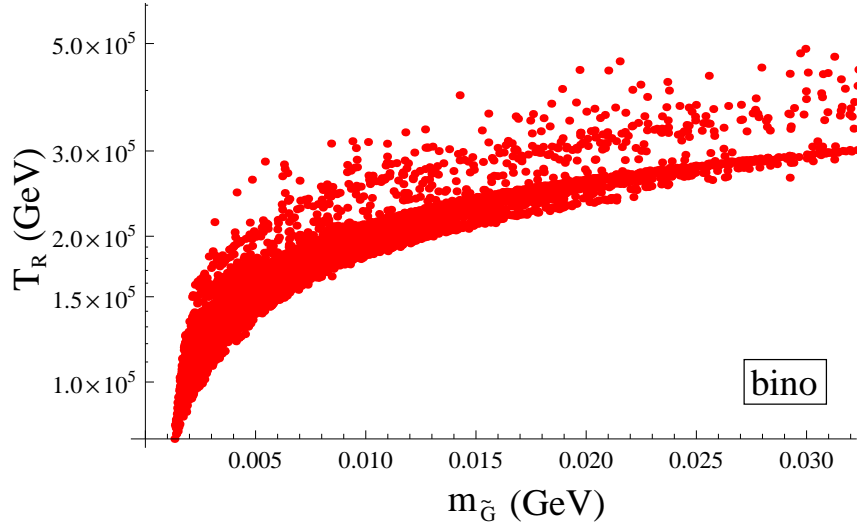}
\end{center}
\caption{Reheating temperature versus gravitino mass (both in GeV) for
the bino case.}
\end{figure}

\begin{figure}
\begin{center}
\includegraphics{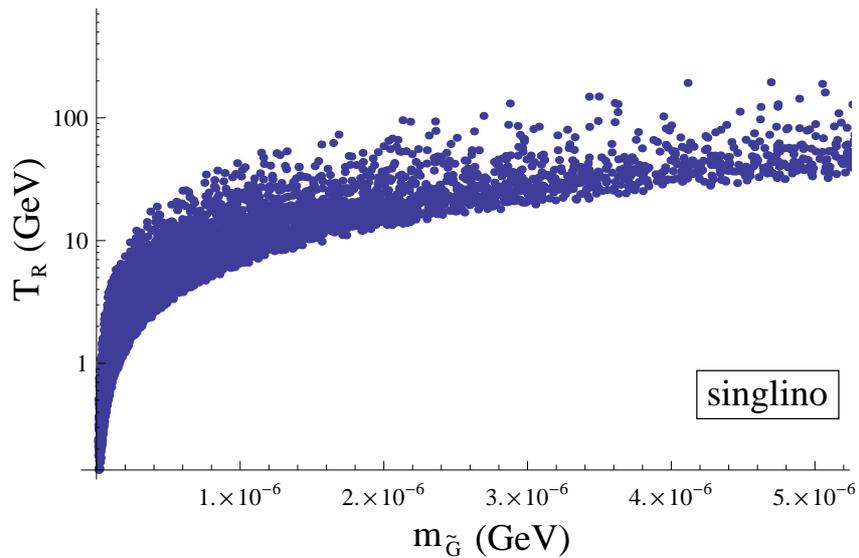}
\end{center}
\caption{Reheating temperature versus gravitino mass (both in GeV) for
the singlino case.}
\end{figure}

\section{Conclusion}

In the framework of the cNMSSM, we have considered a
possible cosmological scenario with the gravitino LSP and the
neutralino NLSP. The gravitino is stable and plays the role of cold
dark matter in the universe, while the neutralino is unstable and
decays to gravitino. We have taken into account the relevant
gravitino production mechanisms, which are i) the NLSP decay, and
ii) scattering processes from the thermal bath. Our results can be
seen in the figures. We have found that i) the gravitino is
necessarily very light, and ii) the reheating temperature after
inflation is two orders of magnitude lower than the temperature
required for thermal leptogenesis. The singlino scenario must be
excluded, while in the bino case it is hardly possible to have a gravitino
in the gravity-mediated SUSY-breaking scheme.


\ack

We acknowledge financial support from FPA2008-02878, and Generalitat 
Valenciana under the grant PROMETEO/2008/004.



\end{document}